# Protein sequestration versus Hill-type repression in circadian clock models

Jae Kyoung Kim[1*]

[1] Department of Mathematical Sciences, Korea Advanced Institute of Science and Technology, 291 Daehak-ro Yuseong-gu, Daejeon, 305-701, Korea.
[*] jaekkim@kaist.ac.kr

**Abstract:** Circadian (~24hr) clocks are self-sustained endogenous oscillators with which organisms keep track of daily and seasonal time. Circadian clocks frequently rely on interlocked transcriptional-translational feedback loops to generate rhythms that are robust against intrinsic and extrinsic perturbations. To investigate the dynamics and mechanisms of the intracellular feedback loops in circadian clocks, a number of mathematical models have been developed. The majority of the models use Hill functions to describe transcriptional repression in a way that is similar to the Goodwin model. Recently, a new class of models with protein sequestration-based repression has been introduced. Here, we discuss how this new class of models differs dramatically from those based on Hill-type repression in several fundamental aspects: conditions for rhythm generation, robust network designs and the periods of coupled oscillators. Consistently, these fundamental properties of circadian clocks also differ among *Neurospora*, *Drosophila*, and mammals depending on their key transcriptional repression mechanisms (Hill-type repression or protein sequestration). Based on both theoretical and experimental studies, this review highlights the importance of careful modeling of transcriptional repression mechanisms in molecular circadian clocks.

## 1. Introduction

We wake up and sleep at the usual times mainly because the level of the hormone melatonin in our brain is elevated and reduced at the right times of day [1]. Our blood pressure also exhibits a daily pattern – it is high in the morning and low at night. Similar daily patterns are observed in other organisms: *Drosophila* eggs hatch only in the morning, and *Neurospora* mold begins producing spores only in the evening. These daily (24hr) rhythms, seen in diverse behavioral, physiological, and developmental processes, are driven by intrinsic self-sustained oscillators, circadian clocks. With these endogenous oscillators, organisms ranging from unicellular bacteria and insects to mammals can anticipate periodic daily changes in the environment, and regulate their cellular activities or behavior to occur at the appropriate times of day and night [2]. Moreover, this intrinsic time-tracking system enables organisms to actively control their physiology in the face of seasonal day-length changes [1]. For instance, mammals can change their sleep phases according to the time of sunset or sunrise [3]. Plants can regulate their starch degradation rate during the night, so that starch reserves are exhausted at sunrise regardless of day-length change [4, 5].

While molecular components underlying circadian clocks vary among organisms, three basic properties are commonly shared, which facilitate the appropriate phase relationships between circadian rhythms and environmental cycles (*e.g.* diurnal cycle) [1, 2]. 1) Rhythms are self-sustained with a period of nearly 24 hr. 2) The period of self-sustained rhythms is maintained over a physiologically relevant temperature range (*i.e.* temperature compensation) [6]. 3) Rhythms can be entrained or reset by external cues such as light or temperature [7, 8]. These dynamic features of circadian rhythms have provided a natural setting for mathematical modeling and led to the publication of more than 600 theoretical studies about circadian rhythms [9-13].

Long before the identification of the molecular basis of circadian clocks, phenomenological models-focusing on the phase and/or amplitude of limit cycle oscillators (*e.g.* Poincare and Van der Pol oscillators) were widely used to study circadian rhythms [7, 14, 15]. Because this approach uses abstract limit cycle oscillators, which are not based on underlying molecular dynamics, the variables and parameters of those models are too abstract to compare with physical quantities. Nonetheless, these abstract models can provide important insight into experimental data on phase relationship of circadian rhythms, such as phase response to external signals [16-18], phase entrainment [19-22], and phase regulation via coupling signal [23-26]. Furthermore, such models have also been used recently to analyse the phase and amplitude information from circadian time-course data [27, 28] and to investigate circadian regulation of other systems (e.g. cell division) [29, 30]. See [10-12] for a detailed review of this type of abstract phase-based models.

Over the last couple of decades, the revolution in molecular experimental techniques has led to the identification of molecular interaction networks underlying circadian clocks in considerable detail (see [2, 31]). In particular, intracellular transcriptional/translational negative feedback loops (NFLs) between activators and repressors have been uncovered as the key oscillatory mechanisms in many organisms including *Neurospora*, *Drosophila* and mammals. These exciting discoveries have spurred the development of molecular-based models in which individual molecular reactions are described by ordinary differential equations [9-13]. Because the typical simulation outputs of such models are time-courses of the rise and fall of specific molecular components, model predictions can be tested directly by experiments. This allows for closer interactions between theories and experiments and enhances our systemic understanding of the molecular basis of circadian clocks. Goldbeter's model of the transcriptional NFL in *Drosophila* was among the first to use this approach [32]. In this model, Hill functions were used to describe transcriptional repression of the NFL following the lead of Goodwin, who was the first person to model oscillations in a simple genetic NFL [33]; see also [34-37]. Since Goodwin's and Goldbeter's pioneering studies, Hill functions have been widely used to model the NFL in circadian clocks of diverse organisms, including *Neurospora* [38-40], *Drosophila* [41-46] and mammals [47-53]. In this review, we refer to this class of models as Hill-type repression based models (HT models). The properties of these models and their contributions to the circadian clock field have been reviewed in [9-13, 35-37].

Recently, a new class of circadian clock models has been developed, which uses protein sequestration-based transcriptional repression rather than Hill-type repression [54-65]. Interestingly, this class of models, which we refer to as protein sequestration-based models (PS models), shows qualitatively different properties from the HT models in several important aspects: conditions for rhythm generation, robust feedback loop designs, and coupling-induced period change. This review describes the differences between these two classes of models, focusing on simple representative examples of each class: the Goodwin model [33-37] and the Kim-Forger model [59]. In addition, we present experimental results that support the conclusion that the properties of circadian clocks also differ among diverse organisms depending on their key repression mechanisms. Overall, the properties of PS models match well with data from *Drosophila* and mammals, while the properties of HT models are consistent with data from *Neurospora*. These differences-depending on the repression mechanisms-indicate that the relevant repression mechanism should be carefully considered in developing models of the circadian clock in specific organisms.

## 2. Two classes of transcriptional NFL models

Although different kinds of molecules are used, a transcription-translation NFL is the common core of the circadian oscillators in diverse organisms, including *Neurospora*, *Drosophila*, and mammals (Fig. 1a and b) (see [2, 31] for details). In the transcriptional NFL, the binding of activator ($A$) to the promoter region of the repressor gene triggers the transcription of repressor mRNA ($M$), which is translated into repressor protein ($C$) in the cytoplasm. Then repressor protein ($R$), after translocation to the nucleus, inhibits the activator and suppresses its own transcription. To describe this transcriptional NFL, many models have taken inspiration from the Goodwin model, which was developed as a hypothetical genetic oscillator long before the molecular components of circadian clocks were identified. The Goodwin model [33, 34] is

$$\dot{M} = \alpha_M f(R) - \beta_M M,$$
$$\dot{C} = \alpha_C M - \beta_C C, \qquad (1)$$
$$\dot{R} = \alpha_R C - \beta_R R,$$

where $\alpha_i$ and $\beta_i$ are production rates and clearance rates of species, respectively. In the Goodwin model, transcriptional repression is described by a Hill function ($f(R)$), which describes how transcriptional activity decreases as repressor concentration ($R$) increases:

$$f(R) = \frac{1}{1 + (R/K_i)^{N_H}}. \qquad (2)$$

The Hill function describes various types of repression mechanisms based on multiple cooperative reactions, such as transcriptional repression via the binding of cooperatively polymerized repressors to the promoter [35, 36, 66, 67]. In this case, the exponent ($N_H$) of the Hill function represents the number of monomers in the polymer, which is rarely large in biological systems [35, 36, 66, 67]. Alternatively, repression based on multiple phosphorylations has been proposed as a more realistic mechanism that can be described by the Hill function with a large Hill exponent [36, 68]. Specifically, when the repressor distributively and cooperatively phosphorylates multiple sites of the activator on a fast timescale, the fraction of activator that is not fully phosphorylated and thus transcriptionally still active is described by a Hill function (Fig. 1c) (see [36, 68] for details). In this case, $N_H$ represents the number of phosphorylation sites on the activator, which can be large, and $K_i$ represents the concentration of phosphatase. Following the Goodwin model (Eqs. 1 and 2), the Hill function has been widely used to describe transcriptional repression in other molecular circadian clock models (HT models) of diverse organisms: *Neurospora* [38-40], *Drosophila* [32, 41-46] and mammals [47-53].

Since the early 2000s, a different transcriptional repression mechanism, based on protein sequestration or protein titration, has been proposed to describe the NFL underlying circadian oscillators [35, 55, 57, 69]. In this case, repressors tightly bind activators to form an inactive 1:1 stoichiometric complex (Fig. 1d). Assuming rapid binding between repressors and activators, the fraction of activators that are not sequestered by the repressors and that are thus transcriptionally active is described by following protein-sequestration function (Fig. 1d) [59, 70-72]:

$$f(R) = \frac{A - R - K_d + \sqrt{(A - R - K_d)^2 + 4AK_d}}{2A} \xrightarrow{K_d \to 0} \begin{cases} 1 - \frac{R}{A} & \frac{R}{A} \leq 1 \\ 0 & \frac{R}{A} > 1 \end{cases} \equiv \left\lfloor 1 - \frac{R}{A} \right\rfloor \quad (3)$$

where $K_d$ is the dissociation constant of the repressor-activator complex. For the case of tight binding (*i.e.* $K_d$ is small), the protein-sequestration function is approximated by a piecewise linear function of the molar ratio between repressors and activators in the nucleus ($\left\lfloor 1 - \frac{R}{A} \right\rfloor$) (Fig. 1d). Specifically, when the molar ratio is greater than 1:1, most activators are sequestered and transcription is almost completely suppressed. On the other hand, as repressor concentration decreases from the 1:1 molar ratio, the released activators account for an approximately linear increase of transcription rate. The approximately piecewise linear curve with the critical point of 1:1 molar ratio (Fig. 1d) is qualitatively different from the sigmoidal curve of the Hill function (Fig. 1c) [61]. The Kim-Forger model modifies the Goodwin model (Eq. 1) by replacing the Hill function (Eq. 2) with the protein-sequestration function (Eq. 3) [59, 61]. Protein sequestration-based repression (Fig. 1c) has also been used in other circadian clock models (PS models) [54-65]. While not discussed in detail in this review, some circadian clock models use a mixture of protein sequestration and Hill-type regulations [73-76].

Although the specific transcriptional repression mechanism is not fully understood in many organisms, protein sequestration (Fig. 1d) appears to be responsible for transcriptional repression in *Drosophila* and mammals [77-83]. Specifically, in these organisms, repressors sequester activators in a 1:1 stoichiometric complex, which inhibits the transcriptional activity of activators. This protein sequestration is the necessary repression step in *Drosophila*, as even after all identified phosphorylation sites are mutated at the activator (CLOCK), the mutated activator is still repressed by the repressor (PER-TIM) [84]. While phosphorylation is not essential for repression in *Drosophila*, it is critical for repression in *Neurospora*: the repressor (FRQ) promotes phosphorylation at multiple sites of the activator (WCC), which prevents WCC from binding to the *frq* gene promoter (Fig. 1c) [85-89]. Furthermore, the transcriptional activity of WCC is not suppressed only by direct complex formation with the FRQ (*i.e.* protein sequestration) [88]. The fact that different organisms employ different mechanisms of transcriptional repression indicates that the repression should be carefully considered when modeling circadian clocks in a specific organism as the properties of models differ dramatically depending on the repression mechanisms described below.

## 3. Conditions for rhythm generation

Under what conditions a circadian clock fails to generate rhythms and how the disrupted rhythms can be restored have been important and fundamental issues. Moreover, these problems are tightly related to human health, as the disruption of circadian rhythms increases the risk of getting various diseases such as insomnia, depression, cancer and diabetes [90]. The essential molecular mechanisms for rhythm generation have been investigated with both HT and PS models. For both the Goodwin model and the Kim-Forger model to generate rhythms, the transcription repression functions (Eqs. 2 and 3) need to show an ultrasensitive response to repressor change in a relative sense at the *steady state* of the models. That is, a large change in relative transcription activity is required for a small change in relative repressor concentration, which can be measured by logarithmic sensitivity ($|dlogf(R)/dlogR|=|(df(R)/dR)(R/f(R))|$). In particular, for both models, it has been shown that the logarithmic sensitivity should be greater than 8 at the steady state (see Appendix for detailed analysis) [34, 59, 91-93]. Importantly, conditions to achieve such high logarithmic sensitivity differ depending on the repression mechanisms (Figs. 1c and d), as described below.

The logarithmic sensitivity of the Hill function is

$$\left|\frac{d\log f(R)}{d\log R}\right| = \left|\frac{R}{f(R)}\frac{df(R)}{dR}\right| = N_H \frac{(R/K_i)^{N_H}}{1+(R/K_i)^{N_H}} \leq N_H \tag{4}$$

which increases as the Hill exponent ($N_H$) or the effective repressor concentration ($R/K_i$) increases (Fig. 2a), and thus the Goodwin model is more likely to generate rhythms with higher amplitudes (Fig. 2b). In particular, since the maximal logarithmic sensitivity cannot be greater than the Hill exponent (Fig. 2a), a Hill exponent greater than 8 is required for the Goodwin model to generate rhythms (Fig. 2b) [34, 91-93]. Since a large Hill exponent is often difficult to achieve in biological systems, various mechanisms to reduce the required Hill exponent have been identified. For instance, the required Hill exponent decreases as more intermediate steps are included to generate time delay in the NFL (Eq. 1), which is known as the secant condition (see Appendix for details) [91-94]. The Michaelis-Menten type of repressor clearance also reduces the necessary Hill exponent as it can serve as an additional source of non-linearity [35, 95-98]. By including such additional mechanisms, the majority of HT models generate rhythms with the Hill exponents of ~4, which are lower than those in the Goodwin model, but still fairly large [40-45, 47-53].

The logarithmic sensitivity of the protein-sequestration function (Eq. 3) is

$$\left| \frac{d\log f(R)}{d\log R} \right| = \frac{R/A}{\sqrt{(1 - R/A - K_d/A)^2 + K_d/A}} \xrightarrow{K_d \to 0} \frac{R/A}{|1 - R/A|}, \tag{5}$$

which increases as the dissociation constant ($K_d$) decreases or the molar ratio between repressor and activator (*R/A*) becomes closer to 1:1 (Fig. 2c) since a sharp transition occurs when the molar ratio is around 1:1 (Fig. 1d) [59, 70, 72]. Consistently, as binding between the repressor and the activator becomes tighter ($K_d \to 0$) or as the average molar ratio throughout a cycle (<*R/A*>) becomes closer to 1:1, the Kim-Forger model generates rhythms with higher amplitudes (Fig. 2d). The importance of the 1:1 molar ratio is also commonly observed in other PS models [55, 57-59]. For instance, in a detailed mammalian PS model, simulated mutant phenotypes with molar ratios far from 1:1 become arrhythmic [59]. Consistently, in a *Drosophila* PS model, all the simulated wild type and rhythmic mutants have molar ratios of about 1:1, though this was not noted in the original work [58]. Similar to the HT models [91-93], time delay via intermediate steps is also important in the PS models. For instance, when an intermediate step for the nuclear translocation of repressor is removed in the NFL (*i.e.* the translated repressor (C) immediately sequesters the activator (A) in Fig. 1a), including the slow binding/unbinding of the activator to the repressor promoter becomes critical to generate rhythms because it can function as an additional intermediate step [69, 99, 100].

Due to the different repression mechanisms, HT and PS models have their own unique requirements to generate rhythms: a large Hill exponent and a 1:1 molar ratio between repressor and activator, respectively (Figs. 2b and d). Consistently, *Neurospora*, *Drosophila* and mammals lose rhythms under different conditions, depending on their key transcriptional repression mechanisms. For instance, as a large Hill exponent is critical for HT models to generate rhythms (Fig. 2b), a large number of phosphorylation sites at the activator (WCC) is required in *Neurospora*. Specifically, as the number of mutated phosphorylation sites increases, the circadian rhythms become weaker and finally arrhythmic [87]. In contrast, even after the mutation of all identified phosphorylation sites at the activator (CLOCK), the circadian clocks of *Drosophila* still generate rhythms [84]. Consistent with PS models (Fig. 2d), the 1:1 molar ratio is critical in the mammalian circadian clocks: as the molar ratio between the repressor (PER1/2) and the activator (BMAL1) becomes closer to 1:1 in mice fibroblasts, the amplitude and sustainability of circadian rhythms are considerably enhanced [77]. Furthermore, the molar ratio is around 1:1 in both the liver tissue of mammals [101] and the S2 cells of *Drosophila* [78]. On the other hand, the molar ratio is much less than 1:1 in the nucleus of *Neurospora* [88] as the molar ratio is not critical for HT models to generate rhythms (Fig. 2b).

## 4. Robust designs of interlocked transcriptional feedback loops

Molecular circadian clocks often contain additional positive feedback loops (PFLs) and/or NFLs regulating activator gene expression (Fig. 3a) on top of the core transcriptional NFL (Fig. 1a) [2, 31]. In the additional NFL, the activator promotes the transcription of *Rev-erbs* (mammals) [102-104], *Vrille* (*Drosophila*) [105-107] or *Csp-1* (*Neurospora*) [108], which represses the expression of the activator gene (Figs. 3a and b). On the other hand, in the additional PFL, the activator promotes the transcription of *Rors* (mammals) [109-112] or *Pdp1* (*Drosophila*) [107], which upregulates the transcription of the activator (Figs. 3a and b). The role of these additional feedback loops was puzzling because theoretically the core transcriptional NFL alone can generate rhythms (Figs. 2b and d). This puzzle has triggered extensive modeling studies to investigate a hypothesis that additional feedback loops enhance the robustness of rhythms.

The additional PFL can generate hysteresis, which provides a time delay in the core NFL [35]. Thus, when the additional PFL is added, the Goodwin model can generate rhythms even with a lower Hill exponent (Fig. 3c) [96]. Similarly, the additional PFL allows other HT models to generate rhythms over a wider range of parameters [35, 46, 113, 114]. This PFL becomes more effective when its timescale is faster than that of the core NFL (*e.g.* the targeting component ($A$) has a shorter half-life than the repressor ($R$)) as it leads to a robust relaxation oscillation (Fig. 3c) [35, 96, 113]. While such a relaxation oscillator based on hysteresis can maintain rhythms with a nearly constant amplitude over a wide range of parameters, its period becomes sensitive and tunable [113, 115]. This raises the question as to whether this function of the additional PFL is beneficial for circadian clocks, whose periods should be robust [115]. In contrast to an additional PFL, an additional NFL in HT models has little effect on the robustness of amplitude and period of oscillations [50]. In fact, an additional NFL often leads to less robust HT models, which generate rhythms over a narrower range of parameters [113]. On the other hand, some studies based on HT models have proposed an advantage of having an additional NFL since it can function as an alternative rhythm generator when the core NFL does not function properly [42, 47, 52]. However, this function as an alternative oscillator needs further experimental validation because the disruption of the core NFL leads to arrhythmic phenotypes in mammals [56, 116], *Drosophila* [117] and *Neurospora* [118, 119].

Similar to HT models, PS models with an additional PFL also have a robust amplitude [55, 57] but a sensitive period [59]. On the other hand, an additional NFL increases the parameter range of rhythm generation with a nearly constant period in the Kim-Forger model and in a detailed mammalian PS model [59]. Specifically, the additional NFL and the core NFL synergistically maintain the 1:1 molar ratio between activator and repressor (Fig. 3d), which is critical for PS models to generate robust rhythms (Fig. 2d). Consistently, the additional NFL also enhances the robustness of period in a *Drosophila* PS model [58]. Interestingly, in contrast to the PFL, as the half-life of the targeting component (*A*) becomes longer, the additional NFL leads to more robust PS models [59].

An additional NFL is critical for the core NFL to generate robust rhythms in PS models [58, 59], but not in HT models [50, 113]. Consistent with these theoretical results, in *Drosophila*, elimination of the additional NFL (*i.e.* cycling *vrille*) results in an arrhythmic phenotype [105]. Due to the mild period phenotype of *Rev-erbα$^{-/-}$* mice [102] and modest rhythmic phenotype of partial *Rev-erbβ* depletion of *Rev-erbα$^{-/-}$* cultured cells [120], *Rev-erbs* have not been considered as core components for robust rhythm generation. However, in a recent study, inducible *Rev-erbα/β* double knockout mice show severely fragmented free-running behavior [103, 104], supporting the critical role of *Rev-erbs* in generating robust rhythms. Furthermore, just as the slower additional NFL is more effective in PS models [59], the half-lives of activators are also considerably longer than those of repressors in mammals [121-125], *Drosophila* [126, 127] and *Neurospora* [76]. Tight regulation of activator level via an additional NFL (Fig. 3d) is also observed [59, 128-130]. While an additional NFL also leads to active regulation of WCC level in *Neurospora*, its elimination (*Csp1-/-*) has little effect on the robustness of rhythms [75, 108]. This is consistent with the prediction of HT models that an additional NFL is not essential for robust rhythms [50, 113] because its active regulation of activator level and thus the 1:1 molar ratio is not important in HT models (Figs. 2a and b).

As the addition of a PFL reduces the robustness of period in both HT and PS models [59, 113, 115], its role in generating robust circadian rhythms has not been observed. In *Neurospora*, an additional transcriptional PFL has not been identified (Fig. 3b). In *Drosophila*, *Pdp1ε* knockdown or overexpression does not alter the circadian oscillation function [131]. *Rorα*, *Rorβ* and *Rorα/γ* mutant mice still show robust free-running with a slight change in period [109-112]. Recent studies show that these additional PFLs appear to function to regulate oscillator output rather than generating robust rhythms: *Pdp1ε* links the circadian clock output to the locomotor activity in *Drosophila* [131], and *Rorγ* plays a role for the circadian regulation of metabolic genes in mammals [110].

## 5. The synchronized periods of coupled oscillators

In mammals, the circadian clocks in peripheral tissues (*i.e.* peripheral clocks) are orchestrated by the master clock, residing in the suprachiasmatic nucleus (SCN) of the hypothalamus [132]. The master clock consists of ~20,000 neurons, each of which generates rhythms with their own periods and phases. Intercellular coupling synchronizes these rhythms, which allows precise timekeeping of the SCN [133-135]. Among various intercellular coupling signals, the most essential one is known to be vasoactive intestinal polypeptide (VIP), which is rhythmically released from a subset of SCN neurons and then promotes the transcription of repressor in other neurons in the SCN [133, 136]. The roles of VIP in the master clock have been widely investigated with mathematical models. Specifically, modeling studies show that for both HT and PS models, VIP signals can synchronize heterogeneous rhythms under various types of couplings, including all-to-all coupling [61, 116, 137, 138], part-to-all coupling [63], local-diffusion coupling [139-142], random coupling [140, 142], and scale-free network coupling [142, 143]. Furthermore, coupling via VIP also enhances the robustness of both classes of models against external or internal perturbations such as entrainment signals [19, 63] and genetic mutations [74, 116]. Recently, the role of another important coupling signal GABA [64, 144] has been investigated with both classes of models [64, 65, 145-147].

While many properties regarding intercellular coupling are commonly shared between HT and PS models, one distinguishing property has been recently reported [61, 148]. When heterogeneous oscillators with different periods are coupled, they can be synchronized with a specific period. This synchronized period with VIP differs considerably by ~3-6 hrs from the population mean period of uncoupled oscillators in many HT models (Fig. 4a left) [61, 137, 140-142]. On the other hand, VIP synchronizes rhythms of PS models with a period similar to their mean period (Fig. 4a right) [61, 63]. This difference regarding synchronized periods can be explained by analysing the phase response curve (PRC) to VIP signal [61, 148]. As VIP promotes repressor gene expression, it can advance or delay the phase of individual oscillators depending on the phase when VIP is given (Fig. 4b). The advance region and delay region of the PRC are similar in the Kim-Forger model, indicating that the coupling signal speeds up and slows down a population of cells in balance. Hence, after coupling, the synchronized period stays near the population mean period of uncoupled cells in the Kim-Forger model. However, the Goodwin model typically has an unbalanced PRC due to the sigmoidal character of the Hill function (Fig. 1c) (see [61] for a detailed analysis).

How does intercellular coupling affect the periods and phases of cells in the SCN? When intercellular coupling is disrupted by enzymatic dispersion (Fig. 4c top) or through the knockout of VIP (Fig. 4c bottom), the standard deviation of periods dramatically increases by 2-to-3 fold [135, 136]. On the other hand, the mean periods of uncoupled cells and of coupled cells show little difference, less than 5% (Fig. 4c), consistent with PS models (Fig. 4a right). In agreement with this feature of the SCN, optogenetic manipulation of the SCN firing rate leads to a balanced PRC in a VIP-dependent manner [149]. PRC responses to VIP have different features depending on the dose of VIP: the PRCs become more unbalanced as dose increases [150].

Because peripheral clocks do not have intercellular coupling, they behave like the uncoupled SCN [116], and their periods are similar to the population mean period of the uncoupled SCN [151]. Thus, when coupling synchronizes a period similar to the population mean of the uncoupled SCN, the periods of the master clock can be kept similar to the periods of peripheral clocks (Fig. 4c). This helps the master clock to orchestrate and synchronize peripheral clocks, which are less likely to be entrained by the master clock as their period difference increases. To achieve this advantageous property for the master clock in mammals, a transition from phosphorylation-based repression in *Neurospora* to protein sequestration appears to be essential, according to observed differences between HT and PS models (Fig. 4a) [61, 148].

Note that models based on phosphorylation-based repression can also have a synchronized period that is similar to the population mean if a different type of intercellular coupling is used, such as sharing a common enzyme for phosphorylation [152]. This type of coupling via sharing a common molecule is possible in *Neurospora* due to incomplete cross walls between cells in most strains. This may explain how the fused strains of *Neurospora* circadian clocks synchronize rhythms with their mean period [153].

## 6. Conclusion

In this review, we compare two classes of circadian clock models, which are based on different transcriptional repression mechanisms: Hill-type repression and protein sequestration-based repression (Fig. 5). This difference of repression mechanisms alone leads to dramatic differences in fundamental properties of the models, such as the conditions for autonomous rhythm generation, the robust transcriptional feedback loop designs, and the synchronized periods induced by intercellular coupling (see Fig. 5 for details). Surprisingly, these properties of "HT models" and "PS models" are in good agreement with experimental data from the circadian clocks of *Neurospora*, and of *Drosophila* and mammals, respectively.

While models with a simple phosphorylation-based repression or protein sequestration-based repression (Figs. 1c and d) successfully capture key properties of the circadian clocks in specific organisms (Fig. 5), the actual repression mechanism appears to be more complex. For instance, in *Neurospora*, FRQ not only inhibits WCC via phosphorylation but also triggers the clearance of WCC from the nucleus by forming an FRQ-WCC complex [76]. In mammals, mechanisms for the displacement of the activator from the gene promoter have not been confirmed, which leaves open the possibility of additional repression mechanisms [83]. Furthermore, to repress BMAL-CLOCK, PER forms a large complex with other molecules [154, 155], which can potentially lead a Hill-type repression via cooperative multi-subunit complex formation. Molecular detail of the repression mechanisms, which could be a combination of phosphorylation and protein sequestration, should be investigated in future experimental studies and model-building because the properties of circadian clocks strongly depend on the repression mechanisms (Fig. 5).

For properties of circadian clocks other than those considered in this review (Fig. 5), the major difference between HT and PS models has not been reported or investigated, to our knowledge. Future work can also investigate whether PS models follow the entrainment properties [37, 45, 47, 142, 156-158] or temperature compensation mechanisms [38, 40, 42, 159-166] identified with HT models. Furthermore, stochastic simulations of HT models commonly indicate that circadian clocks can maintain rhythms even with low numbers of molecules [167-170]. An additional PFL, but no additional NFL, enhances the robustness of HT models against the stochasticity [75, 114]. Investigating whether these findings can be generalized to PS models will be interesting future work.

While this review focuses on *Neurospora, Drosophila* and mammals, interlocked transcriptional NFLs and PFLs also exist in the circadian clocks of other organisms [2, 31]. In particular, in plants, the core transcriptional NFL consists of series of transcriptional repressions, which has a similar design as the repressilator or quadripressilator [5, 171-173]. While the detailed transcriptional repression mechanism has not been fully identified, currently models have assumed Hill-type repression [171, 174-177]. Further research should be undertaken to investigate which type of transcriptional repression mechanism is most appropriate for plant circadian clock models.

Besides circadian clocks, there are many other biological oscillators [178] such as segmentation clocks [179, 180], cell cycle oscillators [181-183], p53 oscillators [184, 185], and synthetic oscillators [186, 187]. While Hill-type regulations have frequently been used in models of these biological oscillators [188-196], the critical roles of protein sequestration were also reported [71, 197]. It would be interesting in future work to explore whether diverse biological oscillators show different properties depending on their key repression mechanisms, as presented in this review (Fig. 5).

## 7. Appendix

**Nondimensionalization of the Goodwin model and the Kim-Forger model**

To reduce the number of parameters discussed in Fig. 2, we assumed that clearance rates are the same (*i.e.* $\beta_M = \beta_P = \beta_R = \beta$) and nondimensionalized the Goodwin model and the Kim-Forger model as described in [59]. Specifically, we scaled the variables and time as

$$M = \frac{\alpha_M}{\beta}\underline{M}, \quad C = \frac{\alpha_M \alpha_C}{\beta^2}\underline{C}, \quad R = \frac{\alpha_M \alpha_C \alpha_R}{\beta^3}\underline{R}, \quad t = \frac{1}{\beta}\underline{t}.$$

We also scaled $K_i$ in the Hill function (Eq. 2) and $K_d$ and $A$ in the protein-sequestration function (Eq. 3) as

$$K_i = \frac{\alpha_M \alpha_C \alpha_R}{\beta^3}\underline{K}_i, \quad K_d = \frac{\alpha_M \alpha_C \alpha_R}{\beta^3}\underline{K}_d, \quad A = \frac{\alpha_M \alpha_C \alpha_R}{\beta^3}\underline{A}.$$

With these scalings, Eq. 1 becomes

$$\dot{\underline{M}} = f(\underline{R}) - \underline{M},$$
$$\dot{\underline{C}} = \underline{M} - \underline{C},$$
$$\dot{\underline{R}} = \underline{C} - \underline{R},$$

where

$$f(\underline{R}) = \frac{1}{1 + (\underline{R}/\underline{K}_i)^{N_H}}$$

or

$$f(\underline{R}) = \frac{(\underline{A} - \underline{R} - \underline{K}_d) + \sqrt{(\underline{A} - \underline{R} - \underline{K}_d)^2 + 4\underline{A}\underline{K}_d}}{2\underline{A}}.$$

Note that the nondimensionalized system depends on two non-dimensionalized parameters, $N_H$ and $K_i$ of the Hill function or $A$ and $K_d$ of the protein-sequestration function. This is why we considered only these parameters and simply assumed $\alpha_i=\beta_i=1$ in Fig. 2.

**Secant condition**

Here, we describe the secant condition introduced in Section 3 (see [91-94] for further details). The order of reaction $g(S)$ with respect to $S$ is defined as $d\log(g(S))/d\log(S)$. For instance, if $g(S)=kS^n$, then the order of reaction becomes $n$. For the NFL described in Eq. 1, the necessary condition for instability at the steady state is

$$\left| \frac{\frac{d\log(\alpha_M f(R))}{d\log(R)} \frac{d\log(\alpha_C M)}{d\log(M)} \frac{d\log(\alpha_R C)}{d\log(C)}}{\frac{d\log(\beta_M M)}{d\log(M)} \frac{d\log(\beta_C C)}{d\log(C)} \frac{d\log(\beta_R R)}{d\log(R)}} \right| \geq Sec(\frac{\pi}{3})^3 = 8,$$

which is known as the secant condition due to the secant function of the right-hand side. The numerator of the left-hand side consists of the orders of clearance reactions at the steady state. The denominator consists of the orders of production reactions with respect to prior species in the feedback loop. Since the order of linear reaction is 1, the above secant condition is simplified as

$$\left| \frac{d\log(\alpha_M f(R))}{d\log(R)} \right| \geq Sec(\frac{\pi}{3})^3 = 8.$$

If the clearance rates of all species are equal ($\beta_M=\beta_P=\beta_R=\beta$) as we assumed in Fig. 2b and d, the above condition with strict inequality becomes a sufficient and necessary condition for the instability of the steady state. Therefore, if the logarithmic sensitivities of the Hill function (Eq. 2) or the protein-sequestration function (Eq. 3) are greater than 8, the steady state becomes unstable. It has also been shown that the unstable steady state leads to periodic solutions in the negative feedback loop [198, 199] as seen in Fig. 2b and d.

Note that the secant condition indicates that conditions for rhythm generation depend on the sensitivity of response in a relative sense (*i.e.* the logarithmic sensitivity) rather than the sensitivity of response in an absolute sense. This explains how the approximately piecewise linear function (Fig. 1d), which is not stiff and thus has a low absolute sensitivity, can generate rhythms (Fig. 2d). Furthermore, it is not true that HT models are more likely to generate rhythms as the steady state become closer to the stiffest point of Hill-function (*i.e. R/Ki*=1), where absolute sensitivity, but not relative sensitivity, is the highest (Fig. 2b). Note that the curves of both the Hill-function and the protein-sequestration function have abrupt bends from "decreasing" to "nearly flat" at points where relative sensitivity is high (Fig. 1c and d). This local similarity between the two curves indicates that the Goodwin model and the Kim-Forger model generate rhythms with an equivalent mechanism in the mathematical sense. However, the steady states of the two models are located at the points of the curves, where the relative sensitivity is high, under different biological conditions as seen in Fig. 2a and c.

For the NFL models (Eq. 3) with *n* intermediate steps, the right-hand side of the secant condition becomes $Sec(\pi/n)^n$, which decreases as *n* increases. For instance, as *n*=3 increases to *n*=4, $Sec(\pi/n)^n$ decreases from 8 to 4. Therefore, as more intermediate steps are included, which leads to more time delay, the secant condition becomes less restrictive, and lower logarithmic sensitivities of Hill functions or protein-sequestration functions are required to generate rhythms.

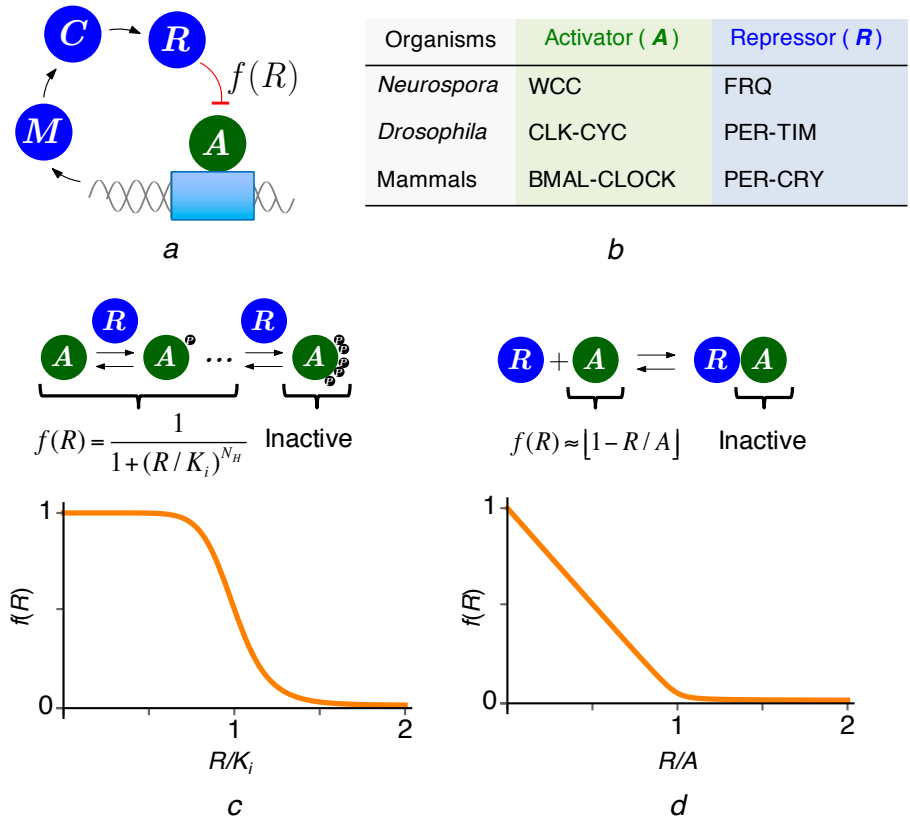

***Fig. 1.** Repression mechanisms closing the core transcriptional NFL of circadian clocks.*
a In the core transcriptional NFL of circadian clocks, the activator ($A$) binding to the promoter of repressor genes leads to the transcription of the repressor *mRNA* ($M$) and then the translation of the repressor protein ($C$) in the cytoplasm. The nuclear translocated repressor protein ($R$) suppresses the transcriptional activity of the activator, which is described with the function $f(R)$ in the Goodwin model and the Kim-Forger model (Eqs. 1-3).
b The list of the activator and repressor proteins in the circadian clocks of *Neurospora*, *Drosophila* and mammals.
c With the phosphorylation-based repression mechanism, the repressor inhibits the activator by triggering phosphorylation at multiple sites of the activator. If the phosphorylation occurs in a distributive and cooperative manner on a fast time-scale, the fraction of the transcriptionally active activator that is not fully phosphorylated is described with the sigmoidal Hill function of the effective repressor concentration ($R/K_i$) (Eq. 2). $N_H$ and $K_i$ represent the number of phosphorylation sites and the concentration of phosphatase, respectively.
d With the protein sequestration-based repression mechanism, the repressor inhibits the activator via forming the 1:1 stoichiometric complex. If the binding and unbinding occurs with a high affinity on a fast time-scale, the fraction of the free activator, which is transcriptionally active, is the approximate piecewise linear function of the molar ratio ($R/A$) (Eq. 3).

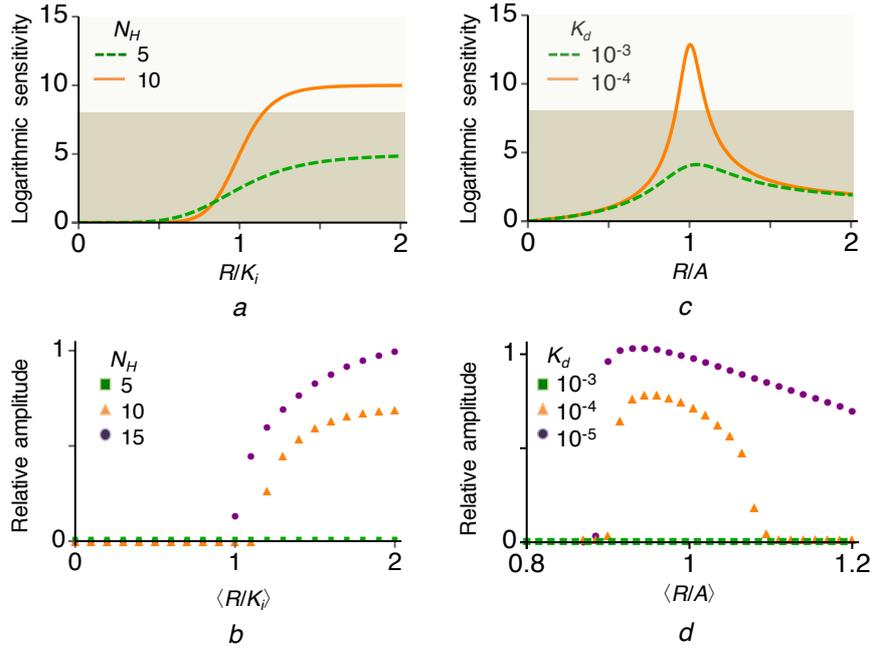

*Fig. 2. Conditions for autonomous rhythm generation differ depending on the transcriptional repression mechanisms.*
a As the Hill exponent ($N_H$) or the effective repressor concentration ($R/K_i$) increases, the logarithmic sensitivity ($dlogf(R)/dlogR$) of the Hill-function (Eq. 4) increases. If the logarithmic sensitivity is less than 8, which is represented as a darker region, the Goodwin model cannot generate rhythms due to the lack of ultrasensitive response in a relative sense (see Appendix for detailed analysis).

b As the $N_H$ or the average of effective repressor concentration ($\langle R/K_i \rangle$) increases, the Goodwin model generates rhythms with higher amplitudes. $\langle R/K_i \rangle = \int_0^\tau R/K_i \, dt$, where $\tau$ is the period of the simulated limit cycle. Here $K_i$ is changed to perturb $\langle R/K_i \rangle$, and $\alpha_i = \beta_i = 1$ is assumed in Eq. 1 (see Appendix for the rationale underlying this assumption).

c As the molar ratio between repressor and activator ($R/A$) becomes closer to 1:1 or the dissociation constant ($K_d$) decreases, the logarithmic sensitivity of the protein sequestration function (Eq. 5) increases. If the logarithmic sensitivity is less than 8, which is represented as a darker region, the Kim-Forger model cannot generate rhythms. Here, $A=0.0659$ (a.u.) and the unit of $K_d$ is the same as that of $A$.

d As the average molar ratio between repressor and activator ($\langle R/A \rangle$) becomes closer to 1:1 or the $K_d$ decreases, the Kim-Forger model generates rhythms with higher amplitudes. Here $A$ is changed to perturb $\langle R/A \rangle$, and $\alpha_i = \beta_i = 1$ is assumed in Eq. 1 (see Appendix for the rationale underlying this assumption).

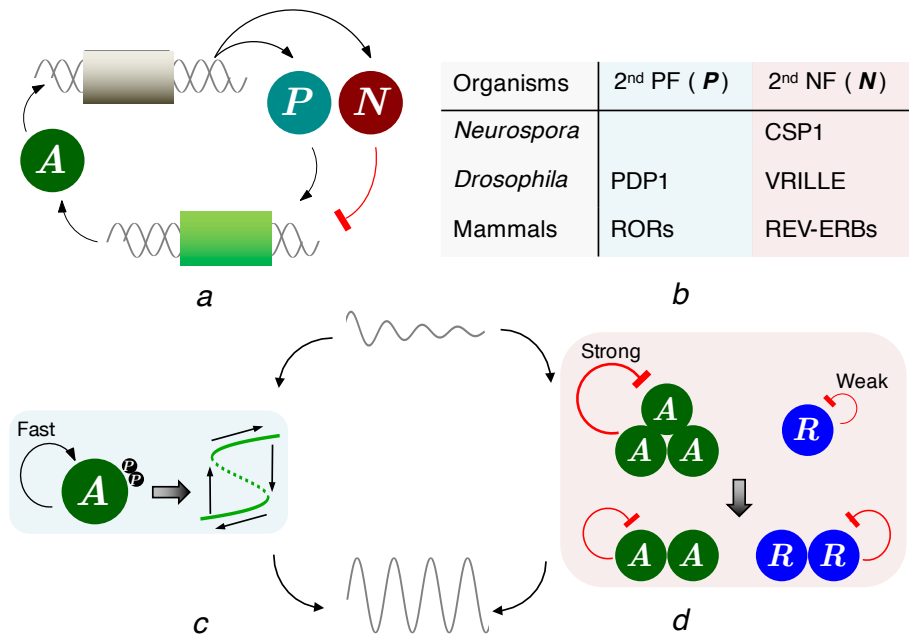

***Fig. 3.*** *The role of additional feedback loops in generating robust rhythms differs depending on the repression mechanisms.*
a Molecular circadian clocks have additional PFLs and/or NFLs with which the activator promotes and suppresses its own gene expression, respectively.
b The mediator proteins of the additional PFLs and NFLs in the circadian clocks of *Neurospora*, *Drosophila* and mammals.
c The fast additional PFL generates a relaxation oscillation based on hysteresis, which reduces the required Hill exponent (*i.e.* the number of phosphorylation sites) for HT models to generate rhythms and enhances the robustness of amplitude.
d The additional NFL and the core NFL synergistically regulate the molar ratio. For instance, when the molar ratio is perturbed to less than 1:1 (*i.e.* excess of activator), the additional NFL strongly suppresses the activator expression, but the core NFL weakly suppresses the repressor expression. This restores the 1:1 molar ratio and enables PS models to sustain rhythms with a robust period.

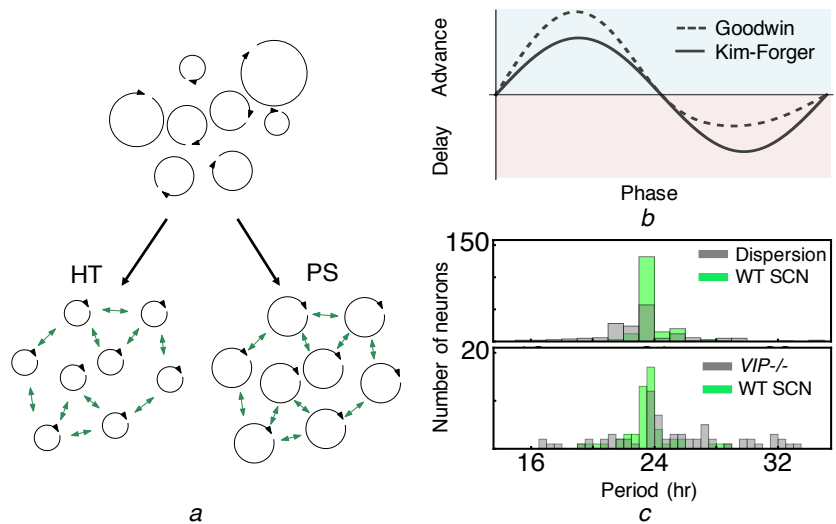

***Fig. 4.** The synchronized periods have different relationships with the mean periods of uncoupled cells depending on the repression mechanisms.*

a The master circadian clock of mammals consists of heterogeneous individual oscillators with different periods and phases, which are represented with the different sizes of circles and the positions of arrows, respectively (top). When these cells are coupled via VIP, in many HT models, synchronized periods are considerably different from the mean periods of the uncoupled oscillators (bottom left). On the other hand, the synchronized periods of the PS models are similar to the population mean periods (bottom right).

b The phase response curve of the Kim-Forger model to VIP has balanced advance and delay regions due to the linear characteristic of the protein-sequestration function (Fig. 1d). In contrast, the Goodwin model has the phase response curve with the unbalanced advance and delay regions due to the sigmoidal characteristic of the Hill function (Fig. 1c).

c When intercellular coupling is disrupted with either enzymatic dispersion or VIP-/-, the distributions of periods become broader among individual cells, but the mean periods show little change. (top) WT SCN: 23.3+/-1 and dispersed SCN: 22.7+/-2.9. (bottom) WT SCN: 23.6+/-1.7 and VIP-/-: 25+/-4. The top panel and bottom panel are reproduced from Ono et al. [135] and Aton et al. [136], respectively, with permission from Nature Publishing Group Ltd.

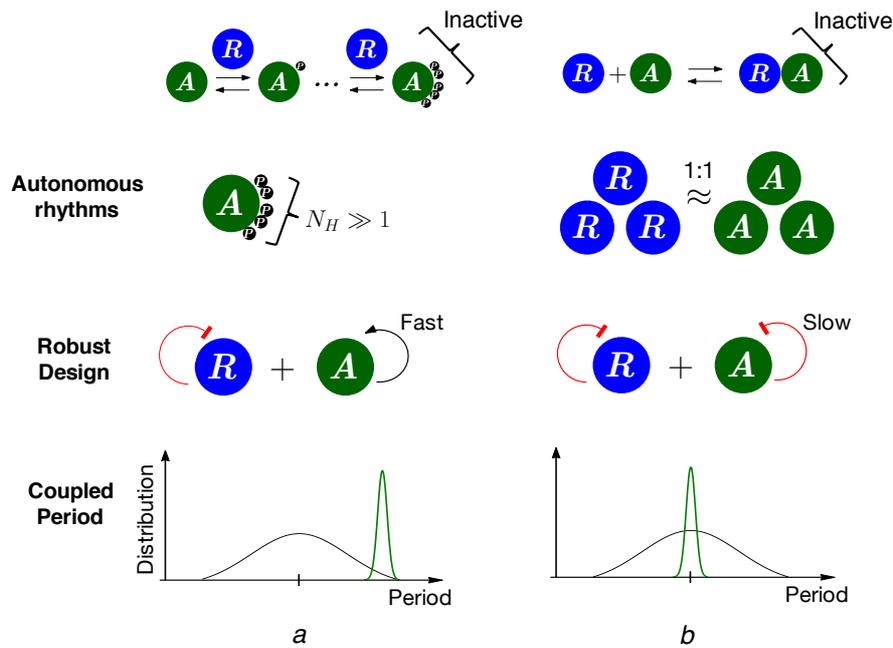

*Fig. 5.* *Diverse properties of circadian clock models differ dramatically depending on the repression mechanisms.*
a When phosphorylation-based repression is used, a large number of phosphorylation sites at activators (*i.e.* a large Hill exponent) is usually required for models to generate rhythms. Furthermore, with a fast additional PFL regulating activator level, but no additional NFL, the core NFL can generate rhythms with a robust amplitude. When individual oscillators are coupled with excitatory signals (*e.g.* VIP in SCN), the synchronized periods of coupled cells (green curves) and the mean periods of uncoupled cells (black curves) are considerably different in many HT models.
b When protein sequestration-based repression is used, models are more likely to generate rhythms when the average molar ratio between repressor and activator becomes closer to 1:1. Furthermore, when a slow additional NFL is included to regulate activator level and thus the molar ratio, the core NFL can generate rhythms with a robust period. The synchronized periods of coupled cells are similar to the mean periods of uncoupled cells in PS models.

# 8. Acknowledgments

We thank Daniel B. Forger, Matthew Bennett, Kresimir Josic, and John Tyson for valuable comments. This work was funded by KAIST Research Allowance Grant G04150020 and the TJ Park Science Fellowship of POSCO TJ Park Foundation G01160001.